\documentclass[11pt,a4paper]{article}
\pdfoutput=1
\usepackage{hyperref}
\usepackage{amssymb,amsmath,amsfonts, mathtools, mathrsfs}
\usepackage[utf8]{inputenc}
\usepackage[dvipsnames]{xcolor}
\usepackage{graphicx}
\usepackage{caption}
\usepackage{tensor}
\usepackage{subcaption}
\usepackage{enumerate}
\usepackage{dsfont}
\usepackage{comment}

\usepackage{verbatim}
\usepackage{amsfonts,euscript,amssymb,braket}
\usepackage{tikz}
\usetikzlibrary{arrows,decorations.markings,patterns}
\usepackage{slashed}

\def\clock{{\count0=\time
		\divide\count0 60
		\ifnum\count0<10 0\fi\the\count0
		\multiply\count0 -60 \advance\count0 \time
		:\ifnum\count0<10 0\fi \the\count0
}}
\newcommand{\timestamp}{{\small\vbox{\hbox{\tt\jobname.tex}
			\hbox{\the\day/\the\month/\the\year, \clock}}}}

\makeatletter
\let\old@startsection=\@startsection
\let\oldl@section=\l@section
\renewcommand{\@startsection}[6]{\old@startsection{#1}{#2}{#3}{#4}{#5}{#6\mathversion{bold}}}
\renewcommand{\l@section}[2]{\oldl@section{\mathversion{bold}#1}{#2}}
\makeatother

\numberwithin{equation}{section}

\begin{document}
	\renewcommand{\thefootnote}{\arabic{footnote}}

	\overfullrule=0pt
	\parskip=2pt
	\parindent=12pt
	\headheight=0in \headsep=0in \topmargin=0in \oddsidemargin=0in

	\vspace{ -3cm} \thispagestyle{empty} \vspace{-1cm}
	\begin{flushright} 
		\footnotesize
		\textcolor{red}{\phantom{print-report}}
	\end{flushright}

\begin{center}
	\vspace{1.2cm}
	{\Large\bf \mathversion{bold} Entanglement entropy and the first law at third order for boosted black branes}	
	
	\vspace{0.8cm} {
		Sabyasachi Maulik$^{\,a,b\,}$\footnote[1]{sabyasachi.maulik@saha.ac.in} and Harvendra Singh$^{\,a,b\,}$\footnote[2]{h.singh@saha.ac.in}
	}
	\vskip  0.7cm
	
	\small
	{\em
		$^{a}\,$Theory Division, Saha Institute of Nuclear Physics, 1/AF, Bidhannagar,\\ West Bengal 700064, India
		\vskip 0.05cm
		$^{b}\,$Homi Bhabha National Institute, Anushakti Nagar, Mumbai,\\ Maharashtra 400094, India
	}
	\normalsize
	
\end{center}

\vspace{0.3cm}
\begin{abstract} 
Gauge/gravity duality relates an AdS black hole with uniform boost with a boosted strongly-coupled CFT at finite temperature. We study the perturbative change in holographic entanglement entropy for strip sub-region in such gravity solutions up to third order and try to formulate a first law of entanglement thermodynamics including higher order corrections. The first law receives important contribution from an entanglement chemical potential in presence of boost. We find that suitable modifications to the entanglement temperature and entanglement chemical potential are required to account for higher order corrections. The results can be extended to non-conformal cases and AdS plane wave background.
\end{abstract}

\vspace{0.3 in}
\section{Introduction}\label{intro}
Of late, studies on quantum information theoretic properties of a strongly coupled QFT using tools from AdS/CFT correspondence \cite{Maldacena:1997re, Gubser:1998bc, Witten:1998qj} have received widespread attention. A pioneering work in this course was the holographic formula for computing entanglement entropy \cite{Ryu:2006bv, Ryu:2006ef} and its covariant generalization \cite{Hubeny:2007xt}. The holographic entanglement entropy (HEE) formula inspired the idea that emergence of space-time in the gravity picture is somehow related to the quantum entanglement degrees of freedom in the dual field theory \cite{VanRaamsdonk:2010pw}. An important observation coming out from the holographic calculations is that there exists a relationship between entanglement entropy of an excited state and the corresponding changes in energy and other macroscopic variables that is reminiscent of the first law of thermodynamics
\begin{equation}\label{law1}
	T_E\Delta S_E = \Delta E - \frac{d-1}{d+1}V\Delta P \,,
\end{equation}
where the \emph{entanglement temperature}, $T_E$ is known to obey a universal behaviour inversely proportional to the size of the entangling region. The \emph{first law of entanglement thermodynamics} was initially established by considering very small excitation over the dual space-time geometry (e.g. pure AdS) \cite{Bhattacharya:2012mi, Allahbakhshi:2013rda, Blanco:2013joa} and also from a CFT point of view \cite {Wong:2013gua}. In fact, in the later work it was argued that any conserved charge associated with the excitation could appear in the law with its conjugate \emph{entanglement chemical potential}, $\mu_E$. Over the years, extensive studies have been made on multiple aspects of the first law of entanglement thermodynamics \cite{Pang:2013lpa, He:2014lfa, Chakraborty:2014lfa, Mishra:2015cpa, Ghosh:2016fop, Bhattacharya:2017gzt, Lokhande:2017jik, Maulik:2019qup, Fujita:2020qvp}. The first law like relationship also plays a central role in reconstructing Einstein's equations for the dual geometry from boundary CFT data \cite{Lashkari:2013koa, Faulkner:2013ica, Swingle:2014uza, Faulkner:2017tkh}.
\par In most of the works on the first law of entanglement so far, the bulk excitation is considered only at leading order and there is no general consensus on its validity beyond the leading order. We extend upon the work of \cite{Mishra:2015cpa}, where second order corrections to holographic entanglement entropy for a strip sub-system on the boundary of a boosted AdS black hole were studied. This gravity solution is dual to a finite temperature CFT with uniform boost in one of the spatial directions. It was shown in \cite{Mishra:2015cpa} that the first law of entanglement could incorporate the second order contributions, provided certain constituents of the law were appropriately modified. In this article we go one step further and find explicit expressions for third order perturbative corrections to the HEE in the same setup. We then try to include these new contributions to the first law by making required redefinition of our entanglement temperature $(T_E)$ and chemical potential $(\mu_E)$. While doing so, we consider first order variation of all variables instead of their relative change over the ground state, as is common in bulk of the literature. Our work is motivated by the idea that the first law of entanglement is in fact, an exact relationship not limited to first order fluctuations. Indeed, in the large subsystem limit, the EE is proportional to the thermal entropy \cite{Ryu:2006ef} and the entanglement temperature should also, then, have a flow towards the Hawking temperature of the black hole. This is not, however, eminent if we restrict the law \eqref{law1} at leading order. Similar line of thought was also advocated in \cite{Kim:2016jwu, Saha:2019ado}. In addition, a system with boost allows us to explicitly check the influence of the Kaluza-Klein momentum on the first law of entanglement and draw a parallel with the law of thermodynamics where it is known to contribute \cite{Maldacena:2008wh}.
\par The rest of the paper is organized as follows: in section \ref{sec2} we introduce the bulk geometry and write down the area functional for HEE, in section \ref{sec3} we consider the narrow strip limit to compute the integrals perturbatively up to third order. A modified first law involving third order corrections is put forward in section \ref{sec4}. In section \ref{sec5} we discuss the HEE of AdS plane wave geometry by taking a special double limit of the results obtained. In section \ref{sec6} we extend our calculations to any generic non-conformal Dp-brane geometry using some simple substitution and argue for a first law in this case as well. Finally, we conclude in section \ref{sec7} with a brief discussion.

\section{Minimal area functional for boosted AdS black hole}\label{sec2}
For static space-time the HEE can be calculated using the Ryu-Takayanagi formula \cite{Ryu:2006bv, Ryu:2006ef}, which asserts that the entanglement entropy associated with a region $A$ with boundary $\partial A$ on the asymptotic boundary of the $(d+1)$-dimensional space-time is dual to the area of a minimal co-dimension $2$ surface $\gamma_{A}$ in the bulk such that $\partial \gamma_{A} = \partial A$
\begin{equation}\label{HEEdef}
	S_A = \frac{Area(\gamma_{A})}{4G_N},
\end{equation}
where $G_N$ is the Newton's constant in $(d+1)$-dimensions. The space-time of our interest is a boosted AdS black hole described by the line element
\begin{equation}\label{BHmetric}
	ds^2 = \frac{R^2}{z^2}\left(-\frac{f\left(z\right)dt^2}{K\left(z\right)}+K\left(z\right)\left(dy-\omega\right)^2+dx_1^2+\cdots+dx_{d-2}^2+\frac{dz^2}{f\left(z\right)}\right),
\end{equation}
with \begin{equation}
	f\left(z\right) = 1 - \frac{z^d}{z_h^d},~~~~ K\left(z\right) = 1 + \beta^2\gamma^2\frac{z^d}{z_h^d},
\end{equation}
where $z_h$ denotes the event horizon of the black hole and $0\leq \beta\leq 1$ is the boost parameter, while $\gamma = \frac{1}{\sqrt{1-\beta^2}}$. $R$ sets the overall curvature of the space-time. The boost is taken in $y$-direction and the same is compactified on a circle of radius $r_y$. The Kaluza-Klein form \begin{equation}\omega = \beta^{-1}\left(1-\frac{1}{K}\right)dt\,. \end{equation}
We are interested in the entanglement entropy of a subsystem $A: x_1 \in \left[-\frac{\ell}{2}, \frac{\ell}{2}\right]$ and $x_j \in \left[0, L \right]~\left(j = 2, 3, \ldots, d-2\right)$ such that $L \gg \ell$. We will always consider a constant time slice of the metric \eqref{BHmetric}. Then the RT surface is conveniently described by the function $x_1\left(z\right)$ and its area is given by the integral
\begin{equation}\label{Aintegral}
	\mathcal{A}_\gamma = 2R^{d-1}V_{d-2}\int_{\epsilon}^{z_*}\frac{dz}{z^{d-1}}\sqrt{K}\sqrt{\frac{1}{f}+x_1^{\prime}(z)^2}\,,
\end{equation}
where $V_{d-2} = 2\pi r_y L^{d-3}$ is the overall volume of the extended directions and the y-circle on the boundary and $\epsilon$ is a UV cut-off put to protect the integral from divergence as $z \to 0$. For ease of calculation we will often assume $R = 1$. Upon extremization this area functional gives a first integral of motion
\begin{equation}
	\frac{x_1^{\prime}(z)\sqrt{K}}{z^{d-1}\sqrt{\frac{1}{f}+x_1^{\prime}(z)^2}} = b\,,
\end{equation}
the constant b may be related to the \emph{turning point} $z_*$ by
\begin{equation*}
	K\left(z_*\right) - b^2z_*^{2d-2} = 0
\end{equation*}
which leads to an integral relation between $z_*$ and the strip-width $\ell$
\begin{equation}\label{turnpt}
	\frac{\ell}{2} = z_* \int_{0}^{1}\frac{dy\,y^{d-1}}{\sqrt{f}\sqrt{\frac{K}{K_*} - y^{2d-2}}},
\end{equation}
where $K_* = K(z_*)$ and $y \equiv \frac{z}{z_*}$. We obtain by substitution in equation \eqref{Aintegral} that
\begin{equation}\label{RTintegral}
	\mathcal{A}_\gamma = \frac{2 V_{d-2}}{z_*^{d-2}}\int_{\frac{\epsilon}{z_*}}^{1} \frac{dy}{y^{d-1}}\sqrt{\frac{K}{f\left(1-\frac{K_*}{K}y^{2d-2}\right)} }
\end{equation}

\section{Perturbative calculation of entanglement entropy}\label{sec3}
Equation \eqref{RTintegral} is unusually hard to solve analytically. In fact, the only known exact solution except pure AdS was found for the $(2+1)$-dimensional BTZ black hole \cite{Hubeny:2007xt}. However in certain regime of the solution space the problem can be addressed.
\par Let us assume $z_* \ll z_h$, in this limit the integrals \eqref{turnpt} and \eqref{RTintegral} may be expressed as a power series in $\frac{z_*^d}{z_h^d}$. If one considers $\mathscr{O}\left(\frac{z_*^d}{z_h^d}\right)$ deviations only then the equation \eqref{turnpt} can be recast as
\begin{equation}\label{turnpt1}
	\frac{\ell}{2} = z_{*}\int_0^1 \frac{dy\,y^{d-1}}{\sqrt{1-y^{2d-2}}}\left(1 + \frac{z_*^d}{z_h^d}\left(\frac{y^d}{2}+\frac{k}{2}\frac{1-y^d}{1-y^{2d-2}}\right)+\mathscr{O}\left(\frac{z_*^{2d}}{z_h^{2d}}\right) \right),
\end{equation}
where we chose to denote $\beta^2\gamma^2$ by $k$, the integral can be solved in terms of beta functions, let us define
\[\int_{0}^{1}\frac{dy\,y^{nd-1}}{\sqrt{1-y^{2d-2}}} = \frac{B\left(\frac{nd}{2d-2},\frac{1}{2}\right)}{2d-2}\equiv b_{n-1}\,.\]
We may then express \eqref{turnpt1} as
\begin{equation}\label{turnpt2}
\frac{\ell}{2} = z_{*} \left[b_0 + \frac{z_*^d}{z_h^d}\left(z_{10} + k\,z_{11}\right)\right],
\end{equation}
where the coefficients are
\begin{align}
	z_{10} &= \frac{b_1}{2},\nonumber \\
	z_{11} &= \left(\frac{d+1}{d-1}\right)\frac{b_1}{2}-\frac{1}{d-1}\frac{b_0}{2}\,.\label{z11}
\end{align}
To write the area of the RT surface in terms of the subsystem length $(\ell)$, we need to invert \eqref{turnpt2}. If we denote the turning point for pure AdS by $\bar{z}_* \equiv \frac{\ell}{2b_0}$ then an approximate relationship at leading order is
\begin{equation}\label{turnpt3}
	z_* = \frac{\bar{z}_*}{1+\frac{\bar{z}_*^2}{z_h^d} \left(\frac{z_{10} + k\,z_{11}}{b_0}\right)+\mathscr{O}\left(\frac{\bar{z}_*^{2d}}{z_h^{2d}}\right)}.
\end{equation}
In a similar vein we can perform a series expansion of the area integral \eqref{RTintegral}
\begin{equation}
	\mathcal{A}_\gamma = \frac{2V_{d-2}}{z_*^{d-2}}\int_{\frac{\epsilon}{z_*}}^1 \frac{dy}{y^{d-1}\sqrt{1-y^{2d-2}}}\left(1 + \frac{z_*^d}{z_h^d}\left( \frac{1+k}{2}y^d + \frac{k}{2}\frac{y^{2d-2}\left(1-y^d\right)}{1-y^{2d-2}} \right) + \mathscr{O}\left(\frac{z_*^{2d}}{z_h^{2d}}\right) \right),
\end{equation}
again all integrations here finally boil down to some beta functions. The ground state (no black hole) result can be very easily read off
\begin{align}
	\mathcal{A}_0 &= \frac{2V_{d-2}}{\bar{z}_*^{d-2}}\int_{\frac{\epsilon}{z_*}}^1 \frac{dy}{y^{d-1}\sqrt{1-y^{2d-2}}}\,, \nonumber \\
	&= \frac{2V_{d-2}}{d-2}\left(\frac{1}{\epsilon^{d-2}} - \frac{b_0}{\bar{z}_*^{d-2}} \right),
\end{align}
which exhibits the characteristic UV divergence, contributions at higher orders are all finite. Making use of the relationship \eqref{turnpt3} they can be expressed as
\begin{equation}\label{area1}
	\frac{\mathcal{A}_\gamma}{V_{d-2}} = \frac{\mathcal{A}_0}{V_{d-2}} + \frac{\bar{z}_*^2}{z_h^d}\,\frac{(d-1) + (d+1)k}{2}b_1 + \mathscr{O}\left(\frac{\bar{z}_*^d}{z_h^d}\right)\,.
\end{equation}

The calculation can be continued at higher orders in $\frac{\bar{z}_*^d}{z_h^d}$, albeit with increasing level of difficulty. The second order results were calculated in \cite{Mishra:2015cpa}. In this work, we go one step further and seek to evaluate the area up to $\mathscr{O}\left(\frac{\bar{z}_*^{3d}}{z_h^{3d}}\right)$, at this order the approximate relationship between the subsystem length $(\ell)$ and the turning point $(z_*)$ is
\begin{equation}\label{turnpt4}
	z_* = \frac{\bar{z}_*}{1+\frac{\bar{z}_*^d}{z_h^d}\frac{Z_1}{b_0}+\frac{\bar{z}_*^{2d}}{z_h^{2d}} \left(\frac{Z_2}{b_0}-d\,\frac{Z_1^2}{b_0^2}\right)+\frac{\bar{z}_*^{3d}}{z_h^{3d}} \left(\frac{d(3 d+1)}{2} \frac{Z_1^3}{b_0^3}- 3d\,\frac{Z_1 Z_2}{b_0^2}+\frac{Z_3}{b_0}\right)+\mathscr{O}\left(\frac{\bar{z}_*^{4d}}{z_h^{4d}}\right)}\,,
\end{equation}
where $Z_1 = z_{10} + k\,z_{11}$ and the new coefficient $Z_2~\text{and}~Z_3$ are
\begin{align*}
Z_2 &= z_{20} + k\,z_{21} + k^2\,z_{22}\,,\\
Z_3 &= z_{30} + k\,z_{31} + k^2\,z_{32} + k^3\,z_{33}\,, 
\end{align*}
such that
\begin{align}
z_{20} &= \frac{3}{8}b_2,\\
z_{21} &= \frac{1}{4}\left(-\frac{d+1}{d-1}\,b_1 + \frac{2d+1}{d-1}\,b_2\right),\\
z_{22} &= \frac{1}{8}\frac{2d-1}{(d-1)^2}\,b_0 - \frac{1}{4}\left(\frac{d+1}{d-1}\right)^2b_1 + \frac{3}{8}\frac{2d+1}{(d-1)^2}\,b_2,\\
z_{30} &= \frac{5}{16}b_3,\\
z_{31} &= \frac{3}{16}\left(-\frac{2d+1}{d-1}\,b_2 + \frac{3d+1}{d-1}\,b_3\right),\\
z_{32} &= \frac{1}{16}\frac{(3d-1)(d+1)}{(d-1)^2}\,b_1 - \frac{1}{8}\left(\frac{2d+1}{d-1}\right)^2b_2 + \frac{1}{16}\frac{(d+3)(3d+1)}{(d-1)^2}\,b_3,\\
z_{33} &= \frac{1}{48}\Bigg( \frac{-3-2d(4d-5)}{(d-1)^3}\,b_0 + \frac{3(3d-1)(d+1)^2}{(d-1)^3}\,b_1 - 9\frac{1+4d(d+1)}{(d-1)^3}\,b_2 \nonumber \\ &~~~~~~~+ \frac{15 + d(47-d(3d-5))}{(d-1)^3}\,b_3 \Bigg).
\end{align}
The explicit integrals that produce these coefficients are listed in appendix \ref{append1}. The minimal area at this order takes the following form
\begin{equation}\label{areafinal}
	\begin{split}
		\frac{\mathcal{A}_\gamma^{(3)}}{V_{d-2}} = \frac{\mathcal{A}_0}{V_{d-2}}\, +\, &\frac{\bar{z}_*^2}{z_h^d}\frac{I_1b_1}{2} - \frac{\bar{z}_*^{d+2}}{z_h^{2d}}\frac{\left(d+2\right)I_1^2b_1^2 - I_2b_0b_2}{4\left(d-1\right)\left(d+2\right)b_0}\, +\\ &\frac{\bar{z}_*^{2d+2}}{z_h^{3d}}\,\frac{2(d+1)(2d+3)I_1^3b_1^3-6(d+1)I_1I_2b_0b_1b_2+I_3b_0^2b_3}{48(d-1)^2(d+1)b_0^2} ,
	\end{split}
\end{equation}
where,
\begin{align}
	I_1 &= (d-1) + (d+1)k\,, \label{HEEcoeff1} \\
	I_2 &= 3(d-1)^2 + 2(d-1)(2d+1)k + 3(2d+1)k^2\,,\label{HEEcoeff2} \\ 
	I_3 &= 15(d-1)^3 + 9(d-1)^2(3d+1)k + 3(d-1)(3d+1)(d+3)k^2 \nonumber \\ &~~~~ - (d-5)(d+3)(3d+1)k^3\,, \label{HEEcoeff3}
\end{align}
all of which are positive quantities. The expressions correctly reproduce the first and second order results of \cite{Mishra:2015cpa}, while the third order result is new. Let us point out that the absolute sign of the change in area of the RT surface alternates at each order; being positive at leading order, negative at second order and so on. This is consistent with earlier observations and hints towards the existence of a complete, non-perturbative expression.
\par From equation \eqref{areafinal} The change in HEE over pure AdS could be written as
\begin{equation}
	S^{(3)} = \frac{\mathcal{A}_{\gamma}^{(3)}-\mathcal{A}_{0}}{4G_N^{(d+1)}} 
\end{equation}
\subsection*{Numerical evaluation of HEE}
The perturbation method illustrated above should work well if $\frac{\bar{z}_*^d}{z_h^d} \ll 1$ or in other words $\ell \ll z_h$ (very narrow strip width). It is instructive to do a numerical evaluation of the minimal area and compare how close our perturbation series analysis can mimic the behaviour.

To perform the numerics we restrict ourselves to $\left(4 + 1\right)$ dimensions. We first regularize the area integral \eqref{RTintegral} by separating out the divergent piece and write the finite part as
\begin{equation}\label{numeric1}
	\mathcal{A}_{\gamma} = \frac{2}{z_*^{d-2}} \left[\int_{0}^{1}\frac{dy}{y^{d-1}}\left(\frac{K(y)}{\sqrt{f(y)} \sqrt{K(y)- K_*y^{2 d-2}}}-1\right)-\frac{1}{d-2}\right].
\end{equation}
 Similarly we also write down the area integral for ordinary AdS space-time
 \begin{equation}\label{numeric2}
 	\mathcal{A}_{0} = \frac{2}{\bar{z}_*^{d-2}} \left[\int_{0}^{1}\frac{dy}{y^{d-1}}\left(\frac{1}{\sqrt{1- y^{2 d-2}}}-1\right)-\frac{1}{d-2}\right].
 \end{equation}
 We choose a few values for the turning point $z_*$ and obtain the corresponding subsystem lengths $\ell$ from equation \eqref{turnpt}, for the same values we also integrate equations \eqref{numeric1} and \eqref{numeric2} to obtain the area difference $\Delta A = \mathcal{A}_{\gamma} - \mathcal{A}_{0}$. We then plot $\Delta A$ on the y-axis and $\ell$ on the x-axis.
 \begin{figure}[t]
 	\centering
 	\includegraphics[scale=0.8]{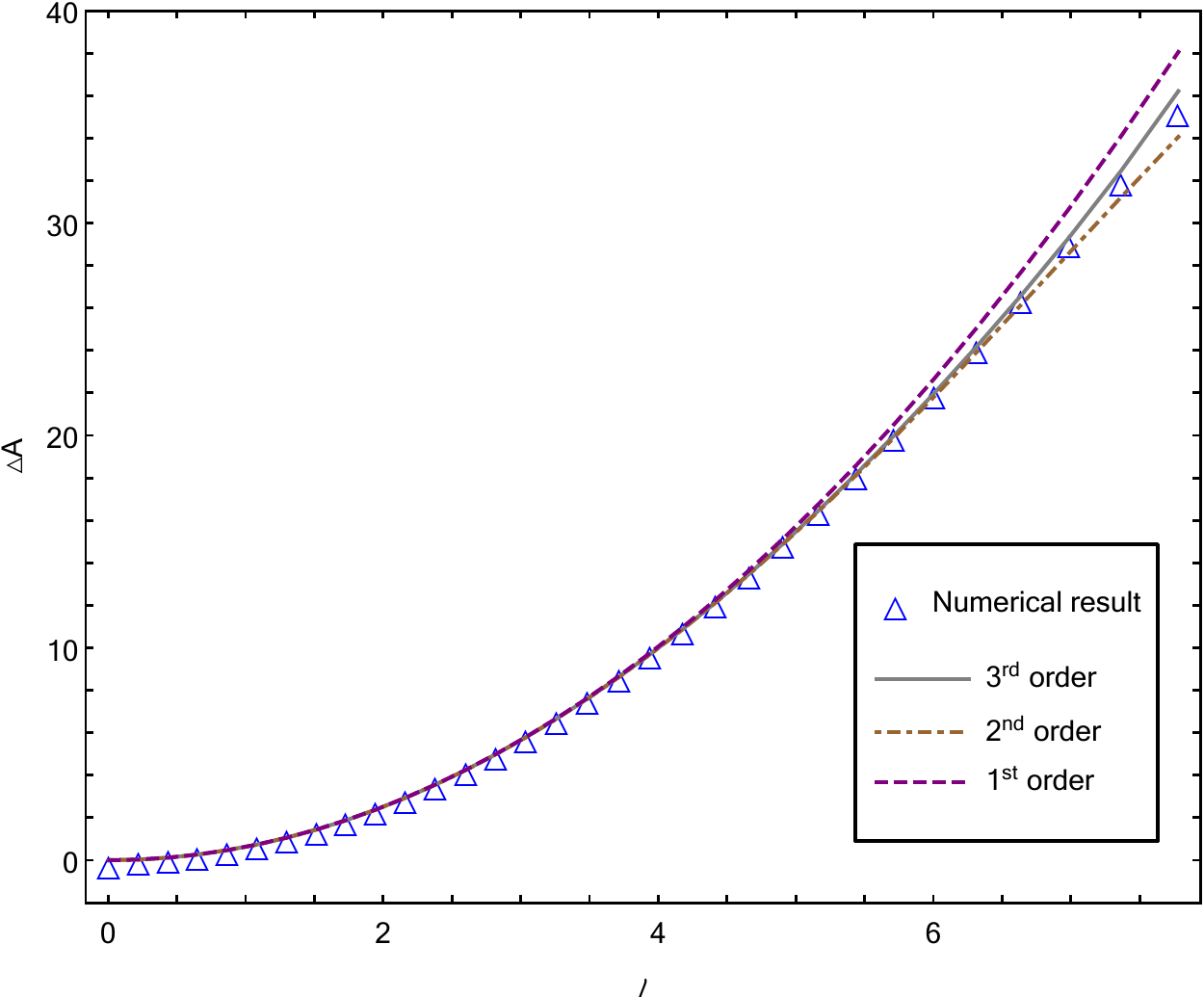}
 	\caption{Area difference of the minimal surface from ground state at different orders for AdS$_5$ and their comparison with numerical result, plot drawn by choosing $\beta = 0.25~\text{and}~z_h = 10$.}
 	\label{fig:area}
 \end{figure}
 \par The results are succinctly expressed in figure \ref{fig:area}, we observe that perturbative results agree to a good extent with numerical values in the region $\ell < z_h$ and the accuracy increases with inclusion of higher order terms.

\section{First law of entanglement thermodynamics}\label{sec4}
The HEE is found to depend on two parameters, viz. the horizon distance $z_h$ and the boost parameter $\beta$ (or equivalently $k$); hence, any infinitesimal change in these two parameters is bound to affect the HEE. Our next task is to express first order variation of $S^{(3)}$ with respect to these parameters as a `first law' like relationship. 
\par To this end, we first require the components of the boundary stress-energy tensor which can be found from a Fefferman-Graham expansion \cite{Balasubramanian:1999re, Kraus:1999di, Bianchi:2001kw} of the metric \eqref{BHmetric}, the expansion leads us to \cite{Mishra:2015cpa}
\begin{equation}
	\langle T_{\mu\nu} \rangle \sim \frac{d}{z_h^d}
	\begin{pmatrix}
	\frac{d-1}{d} + k & \beta\gamma^2 & 0 & \cdots\\
	\beta\gamma^2 & \frac{1}{d} + k & 0 & \cdots\\
	0 & 0 & \frac{1}{d} & \cdots \\
	\vdots & \vdots & \vdots & \ddots
	\end{pmatrix}
\end{equation}
The $T_{01}$ component quantifies the charge due to momentum in the y-direction, we choose an appropriate normalization factor and write the following conserved quantities
\begin{align}
	E &= \frac{V_{d-2}}{16\pi G_N^{d+1}}\left(\frac{d-1}{d} + k\right)\frac{d\,\ell}{z_h^d}\,, \\
	P &= \frac{r_y}{8 G_N^{d+1}}\frac{1}{z_h^d}\,,\\
	N &= \frac{r_yV_{d-2}}{16\pi G_N^{d+1}}\beta\gamma^2\frac{d\,\ell}{z_h^d}\,,
\end{align}
In the above expressions, $E, P \text{~and~} N$ stand for the changes in energy, pressure transverse to the boost and the momentum charge due to boost in $y$-direction, respectively. We also define an \emph{entanglement chemical potential} $\mu_E$ as the value of the Kaluza-Klein gauge field at the turning point \cite{Mishra:2015cpa}
\begin{equation}\label{chempot}
	\mu_E \equiv \frac{1}{r_y\beta}\left(1-\frac{1}{K_*}\right),
\end{equation}
it can be expressed order by order using the perturbative expression \eqref{turnpt4}
\begin{align} \label{pcpotads}
	\mu_E^{(1)} &= \frac{\beta\gamma^2}{r_y}\frac{\bar{z}_*^d}{z_h^d}\,,\nonumber \\
	\mu_E^{(2)} &= \frac{\beta\gamma^2}{r_y}\left(\frac{\bar{z}_*}{z_h}\right)^d\left(1 - \left(\frac{\bar{z}_*}{z_h}\right)^d\left(k + d\frac{Z_1}{b_0}\right)\right).
\end{align}
It is straightforward to write the first law at leading order, the entanglement entropy is
\begin{equation*}
	S^{(1)} = \frac{V_{d-2}}{8G_N^{d+1}}\frac{\bar{z}_*^2}{z_h^d}\,\left((d-1) + (d+1)k\right) b_1
\end{equation*}
from the expressions of $\mu_E$ we make out that any contribution from a $\mu_E\delta N$ like term should not occur before second order. One can easily check that at first order \cite{Bhattacharya:2012mi, Allahbakhshi:2013rda, Mishra:2015cpa}
\begin{equation}
	\delta S^{(1)} = \beta_E^{(1)}\left(\delta E - \frac{d-1}{d+1}V\delta P\right),
\end{equation}
where $V = \ell V_{d-2}$ is the total volume of the subsystem on the boundary. The inverse \emph{entanglement temperature} (at first order) $\beta_E^{(1)}$ is given by
\begin{equation}\label{temp1}
	\beta_E^{(1)} = \left(\frac{d+1}{d}\right)\frac{\pi b_1}{2b_0^2}\ell\,.
\end{equation}

Non-triviality enters at sub-leading orders as it is not transparent how one should take care of subsequent corrections. The authors of \cite{Mishra:2015cpa} argued that higher order corrections could be incorporated in the law through appropriate re-definition of the entanglement temperature, chemical potential and subsystem volume. We take a related but slightly different approach. Let us illustrate with the change in HEE up-to second order; we claim
\begin{equation}
	\delta S^{(2)} = \boldsymbol{\alpha} \left(\delta E - \frac{d-1}{d+1}V\delta P \right) - \boldsymbol{\zeta}\, \mu_E^{(1)}\delta N\,,
\end{equation}
which, after performing the variation, leads to
\begin{equation*}
	\begin{split}
		\frac{V_{d-2}}{8G_N^{(d+1)}}\left[\delta z_h^{-d}\frac{\partial S^{(2)}}{\partial z_h^d} + \delta k \frac{\partial S^{(2)}}{\partial k}  \right] = \delta z_h^{-d}\left(\boldsymbol{\alpha}\left(\frac{\partial E}{\partial z_h^d} - \frac{d-1}{d+1}V \frac{\partial P}{\partial z_h^d}\right) - \boldsymbol{\zeta}\mu_E^{(1)} \frac{\partial N}{\partial z_h^d} \right) + \\ \delta k \left(\boldsymbol{\alpha} \frac{\partial E}{\partial k} - \boldsymbol{\zeta}\mu_E^{(1)}\frac{\partial N}{\partial k} \right).
	\end{split}
\end{equation*}
We refrain from writing full expressions to avoid clutter. To determine the coefficients $\boldsymbol{\alpha}$ and $\boldsymbol{\zeta}$ we may equate the coefficients of $\delta z_h^{-2}$ and $\delta k$ from both sides, this yields the unique solution
\begin{align}
	\boldsymbol{\alpha} &\simeq \beta_E^{(1)}\left(1 + \left(\frac{\bar{z}_*}{z_h}\right)^d\frac{\alpha_1}{\beta_E^{(1)}} + \mathscr{O}\left(\frac{\bar{z}_*^{2d}}{z_h^{2d}}\right)\right),\\
	\boldsymbol{\zeta} &\simeq \frac{\pi b_2}{b_0^2}\frac{d^2 + d - 2 - k \left(2 + 3d -2d^2\right)}{d (d+2) (d - 1 + k(d-3))} \ell + \mathscr{O}\left(\frac{\bar{z}_*^{d}}{z_h^{d}}\right)\,,
\end{align}
where,
\begin{equation}
	\alpha_1 = -\frac{(d+1) \left((d+2)b_1^2-3 b_0 b_2\right)}{d (d+2)b_0} + k\, \frac{(d+1) \left((2d-5)(2d+1)b_0b_2 - \left(d^3-7 d-6\right)b_1^2\right)}{d(d-1)(d+2) ((d-3) k+d-1)b_0}\,.\\
\end{equation}
It is sufficient to determine $\boldsymbol{\zeta}$ at leading order since $\mu_E^{(1)}\delta N$ is itself a second order quantity, we also note that at leading order $\boldsymbol{\alpha} = \beta_E^{(1)}$. It is then appropriate to define $\boldsymbol{\alpha}$ as the inverse entanglement temperature at second order and redefine our entanglement chemical potential as
\begin{align}
	\beta_E^{(2)} &\equiv \boldsymbol{\alpha}\,,\\
	\bar{\mu}_E^{(1)} &\equiv \frac{\boldsymbol{\zeta}}{\boldsymbol{\alpha}} \mu_E^{(1)} \simeq \frac{2\left(d^2 + d - 2 - k \left(2 + 3d -2d^2\right)\right)b_2}{(d+1) (d+2) (d - 1 + k(d-3))b_1}\,\mu_E^{(1)}\,.
\end{align}
The above redefinition allows us to express the variation of HEE in the desired form
\begin{equation}
	\delta S_E^{(2)} = \beta_E^{(2)}\left(\delta E - \frac{d-1}{d+1}V\delta P - \bar{\mu}_E^{(1)}\delta N \right).
\end{equation}
In a similar way we can extend the first law up-to third order, the modified entanglement temperature and chemical potential would be
\begin{align}
	\beta_E^{(3)} &= \beta_E^{(1)}\left(1 + \left(\frac{\bar{z}_*}{z_h}\right)^d\frac{\alpha_1}{\beta_E^{(1)}} + \left(\frac{\bar{z}_*}{z_h}\right)^{2d}\frac{\alpha_2}{\beta_E^{(1)}} + \mathscr{O}\left(\frac{\bar{z}_*^{3d}}{z_h^{3d}}\right) \right), \\
	\bar{\mu}_{E}^{(2)} &= \frac{\boldsymbol{\zeta}}{\boldsymbol{\alpha}}\mu_E^{(2)}\,,
\end{align}
the new coefficient $\alpha_2$ is given by
\begin{equation}
	\alpha_2 = \frac{\alpha_{20} + k\, \alpha_{21} + k^2\, \alpha_{22}}{8d(d-1)^2\left((d-3)k + d - 1\right)b_0^2}
\end{equation}
where,
\begin{equation*}
	\begin{split}
		\alpha_{20} &= \left(15 b_0^2 b_3 - 18(d+1)b_0 b_1 b_2 + 2 (d+1) (2 d+3) b_1^3\right)\left((d-1)^2\left((d-3)k + d - 1\right)\right), \\
		\alpha_{21} &= (d-1)\left((3 d+1) (11 d-15)b_0^2b_3+2 (d (45-4 d (6 d+1))+25)b_0b_1b_2\right.\\ &\left.~~~~+ 2 (d+1)^2 (2 d+3) (3 d-5)b_1^3\right)\,, \\
		\alpha_{22} &= (d+3)(3 d-7)(3 d+1)b_0^2b_3 - 2 (2 d+1) (d+1) ((4d-3)d-19) b_0 b_1 b_2\\ &~~~~+ 2 (2 d+3) (d-3) (d+1)^3 b_1^3\,.
	\end{split}
\end{equation*}
The correct dressing factor for the chemical potential at this order involves first order terms in both $\boldsymbol{\zeta}$ and $\boldsymbol{\alpha}$; it is expressed as
\begin{equation}
	\begin{split}
		\frac{\boldsymbol{\zeta}}{\boldsymbol{\alpha}} = &\frac{2\left(d^2 + d - 2 - k \left(2 + 3d -2d^2\right)\right)b_2}{(d+1) (d+2) (d - 1 + k(d-3))b_1}\\
		&-\left(\frac{\bar{z}_*^{d}}{z_h^{d}}\right)\Big[2(d-2)(d+1)(d+2) k ((d-3) k+d-1) \left(d^2+(d-2)(2d+1)k+d-2 \right)b_1 b_2\\ &+ (d+2)^2 ((d-3) k+d-1) \left(3 d \left(d^2-3\right)+(d-2)(d+3)(3 d+1)k^2\right.\\ &\left.+ 2d(d-1)(3d+1)k + 6\right)b_1 b_3 - 4 (d+1) \left(d^2+(d-2)(2d+1) k+d-2\right)\\ &\left((2 d-5)(2d+1)k^2 + 6(d-1)^2 k + 3(d-1)^2\right)b_2^2\Big]\\ &\left(2 b_1^2 (d-1)(d+1)^2 (d+2)^2 ((d-3)k+d-1)^2\right)^{-1}.
	\end{split}
\end{equation}

\begin{figure}[t]
	\begin{subfigure}{0.48\textwidth}
		\centering
		\includegraphics[width=\textwidth]{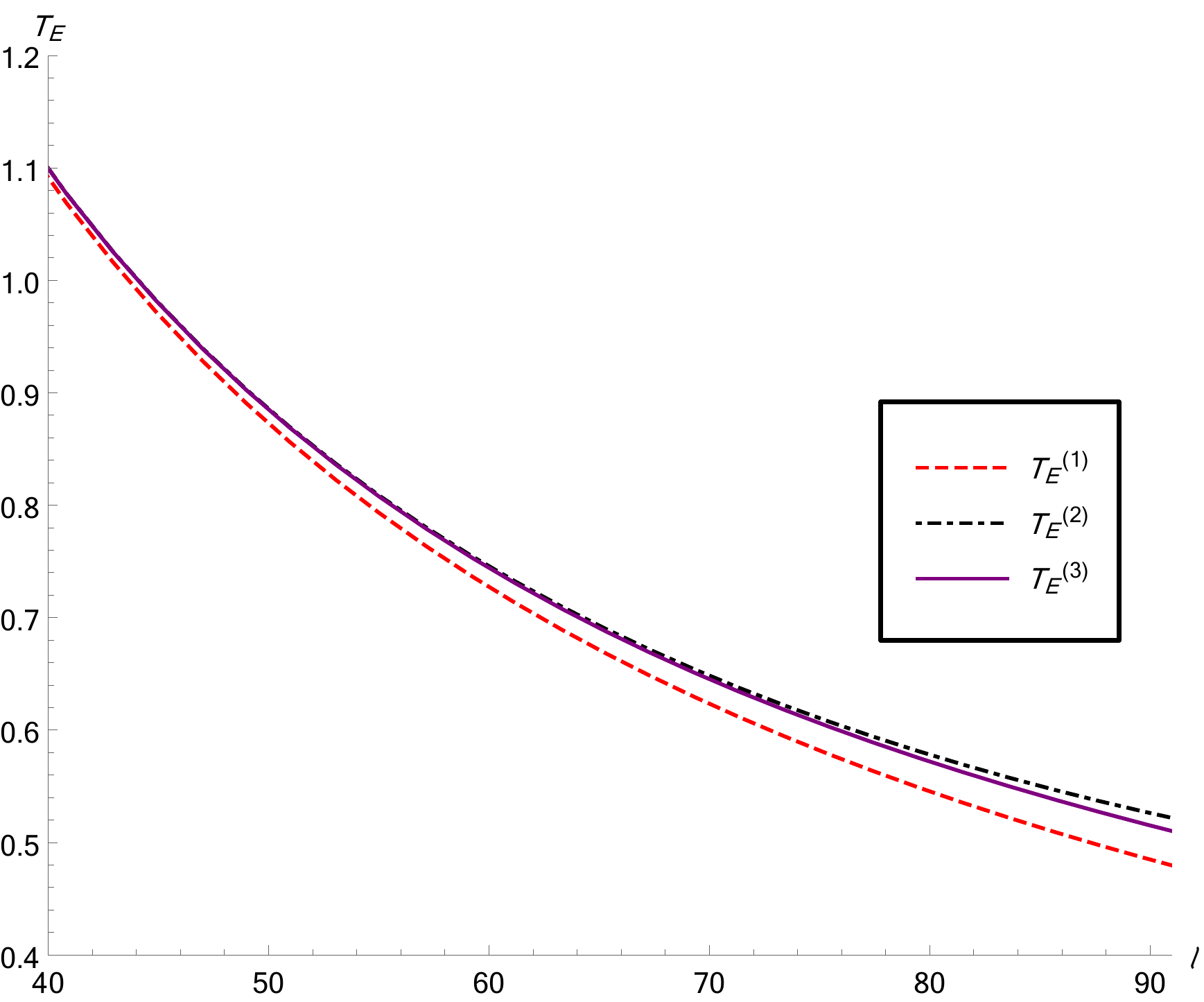}
		\caption{AdS$_4$}
	\end{subfigure}
	\hfill
	\begin{subfigure}{0.48\textwidth}
		\centering
		\includegraphics[width=\textwidth]{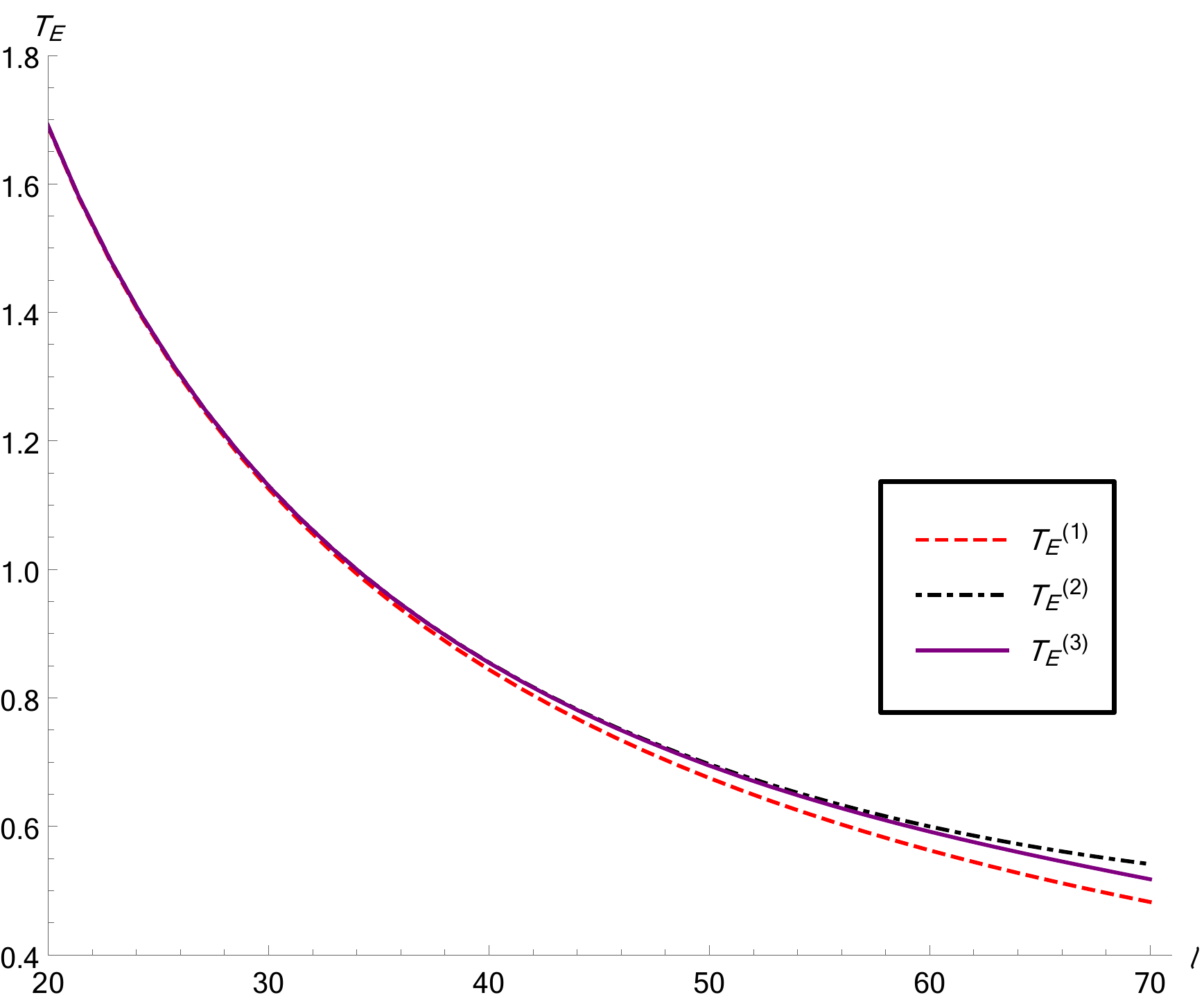}
		\caption{AdS$_5$}
	\end{subfigure}
	\caption{The entanglement temperature with and without higher order corrections for $d=3~\text{and}~4$, plot drawn for $z_h=100$ and $\beta=0.25$.}
	\label{fig:temperature}
\end{figure}

The first law of entanglement including third order corrections is, therefore,
\begin{equation}
\delta S^{(3)} = \beta_E^{(3)} \left(\delta E - \frac{d-1}{d+1}V\delta P  - \bar{\mu}_E^{(2)}\delta N\right), \end{equation}
In figure \ref{fig:temperature} we have shown the dependence of entanglement temperature on subsystem width ($\ell$) with and without higher order corrections. At the leading order $T_E \sim \ell^{-1}$ and it simply decays to zero as $\ell \to \infty$. We note that the decay is somewhat dampened by corrections, the leading order of which is positive. We also note an alternating property of signs as we go on including higher order terms to $T_E$, with the first order being positive, the second order negative and so on. This strongly suggests that the entanglement temperature finally converges to a unique value once all possible corrections are taken care for and matches with the black hole temperature in the large system size limit. However, it is not clear if such a conclusion can be drawn from a simple perturbation series analysis as ours.

\section{Large boost limit and AdS plane wave}\label{sec5}
It is possible to consider a simultaneous limit where the black hole horizon is allowed to shrink while the boost parameter is taken to be very large, we let 
\begin{equation}\label{doublelimit} z_h \to \infty,~~~~\beta \to 1,~~~~\frac{\beta^2\gamma^2}{z_h^d}=\frac{1}{z_I^d}=\text{fixed}. \end{equation}
Such a double limit has been explored before in \cite{Singh:2010zs, Singh:2012un} in the context of non-relativistic holography. In this limit the background geometry \eqref{BHmetric} reduces to an AdS plane wave
\begin{equation}\label{PWmetric}
ds^2 = \frac{R^2}{z^2}\left(-\frac{dt^2}{K}+K\left(dy-\left(1-K^{-1}\right)dt\right)^2+dx_1^2+\cdots+dx_{d-2}^2+dz^2\right),
\end{equation}
where $K = 1 + \frac{z^d}{z_I^d}$. $z_I$ is representative of the energy or momentum of the wave travelling in $y$-direction. The entanglement entropy of AdS plane wave has been studied previously in \cite{Narayan:2012ks, Mukherjee:2014gia, Mishra:2016yor}. We can use our results \eqref{areafinal} and \eqref{HEEcoeff1} - \eqref{HEEcoeff3} and apply the above double limit to obtain the HEE for a strip system in the plane wave background, given that we maintain $\frac{\bar{z}_*^d}{z_I^d}\ll 1$. Up to third order, then, the change in HEE over ground state can be expressed as
\begin{equation}\label{HEEwave}
\begin{split}
S^{(3)}_{\mathrm{plane}} = \frac{V_{d-2}}{8G_N^{(d+1)}}\frac{\bar{z}_*^2}{z_I^d}&\left(I_1b_1 - \frac{\bar{z}_*^{d}}{z_I^{d}}\frac{\left(d+2\right)I_1^2b_1^2 - I_2b_0b_2}{2\left(d-1\right)\left(d+2\right)b_0}\, +\right.\\ &\left.\frac{\bar{z}_*^{2d}}{z_I^{2d}}\,\frac{2(d+1)(2d+3)I_1^3b_1^3-6(d+1)I_1I_2b_0b_1b_2+I_3b_0^2b_3}{24(d-1)^2(d+1)b_0^2}\right) ,
\end{split}
\end{equation}
which is similar in appearance to that for a black hole, except that 
\begin{align*}
I_1 &= (d+1)\,, \\
I_2 &= 3(2d+1)\,, \\
I_3 &= - (d-5)(d+3)(3d+1)\,,
\end{align*}
are the new coefficients obtained by taking the limit \eqref{doublelimit} in eqs. \eqref{HEEcoeff1}-\eqref{HEEcoeff3}.
\par Let us try to figure out the form of the entanglement first law for AdS plane wave. The energy and momentum charge of the CFT excitation dual to this geometry become
\begin{align}
E_{\mathrm{plane}} &= \frac{V_{d-2}}{16\pi G_N^{d+1}}\frac{d\,\ell}{z_I^d}\,, \\
N_{\mathrm{plane}} &= \frac{r_yV_{d-2}}{16\pi G_N^{d+1}}\frac{d\,\ell}{z_I^d}\,,
\end{align}
while the pressure along all $x_i$ directions vanish identically. Similarly, the entanglement chemical potential for the plane wave geometry at different orders can be written by applying the limit \eqref{doublelimit} in eqs. \eqref{pcpotads} 
\begin{align}
\mu_E^{(1)} &= \frac{1}{r_y}\frac{\bar{z}_*^d}{z_I^d}\,,\nonumber \\
\mu_E^{(2)} &= \frac{1}{r_y}\left(\frac{\bar{z}_*}{z_I}\right)^d\left(1 - \left(\frac{\bar{z}_*}{z_I}\right)^d\left(1 + d\frac{z_{11}}{b_0}\right)\right),
\end{align}
with $z_{11}$ being the same as in eqn. \eqref{z11}. Again we see that any contribution to the law from a $\mu_E\delta N_{\mathrm{plane}}$ like term cannot occur at first order where the HEE simplifies to
\begin{equation*}
	S_{\mathrm{plane}}^{(1)} = \frac{V_{d-2}}{8G_N^{(d+1)}}\frac{\bar{z}_*^2}{z_I^d} (d+1)b_1.
\end{equation*}
If we consider first order variation of this expression w.r.t the new scale $z_I$ we can easily establish that
\begin{equation}
	\delta S_{\mathrm{plane}}^{(1)} = \left(\frac{d+1}{d}\right)\frac{\pi b_1}{2b_0^2}\ell\,\delta E_{\mathrm{plane}}.
\end{equation}
Thus the entanglement temperature for AdS plane wave is the same as that for AdS black hole at first order: $\beta_E^{(1)} = \frac{\pi b_1}{2b_0^2}\left(\frac{d+1}{d}\right)\ell$.
\par We should, nevertheless, include the chemical potential when we consider second order or higher corrections because $\mu_E$ is significant from second order onwards. Hence, at third order we are led to adding the following contribution to the first law
\[\mu_E^{(2)}\delta N_{\mathrm{plane}} = \frac{V_{d-2}}{16\pi G_N^{d+1}}d\,\ell \left(\frac{\bar{z}_*}{z_I}\right)^d\left(1 - \left(\frac{\bar{z}_*}{z_I}\right)^d\left(1 + d\frac{z_{11}}{b_0}\right)\right) \delta z_I^{-d}\,. \]
So that the law at third order takes the canonical form
\begin{equation}
	\delta S_{\mathrm{plane}}^{(3)} = \beta_E^{(3)}\left(\delta E_{\mathrm{plane}} - \mu_E^{(2)}\delta N_{\mathrm{plane}} \right),
\end{equation}
with the inverse entanglement temperature at third order, $\beta_E^{(3)}$ being given by
\begin{equation}
\begin{split}
	\frac{\beta_E^{(3)}}{\beta_E^{(1)}} =~ &\frac{\delta E_{plane} - \mu_E^{(2)}\delta N_{plane}}{\beta_E^{(1)} \delta S_{plane}^{(3)}} \\
	\approx~ & 1 - \left(\frac{\bar{z}_*}{z_I}\right)^d \left(1+\frac{b_1^2 (d+1)^2 (d+2)-3 b_0 b_2 (2 d+1)}{b_0 b_1 (d-1) (d+1) (d+2)}\right) \\ &+
	\left(\frac{\bar{z}_*}{z_I}\right)^{2d} \left(\frac{b_0 d-2 b_0+b_1 d^2+b_1 d}{2 b_0 (d-1)} + \frac{\left(b_1^2 (d+1)^2 (d+2)-3 b_0 b_2 (2 d+1)\right)^2}{b_0^2 b_1^2 (d-1)^2 (d+1)^2 (d+2)^2} \right. \\& \left. +\frac{b_1^2 (d+1)^2 (d+2)-3 b_0 b_2 (2 d+1)}{b_0 b_1 (d-1) (d+1) (d+2)} \right. \\& \left. -\frac{b_0^2 b_3 (5-d) (d+3) (3 d+1)-18 b_0 b_1 b_2 (2 d+1) (d+1)^2+2 b_1^3 (2 d+3) (d+1)^4}{8 b_0^2 b_1 (d-1)^2 (d+1)^2}\right)
\end{split}
\end{equation}
\begin{figure}[t]
	\begin{subfigure}{0.48\textwidth}
		\centering
		\includegraphics[width=\textwidth]{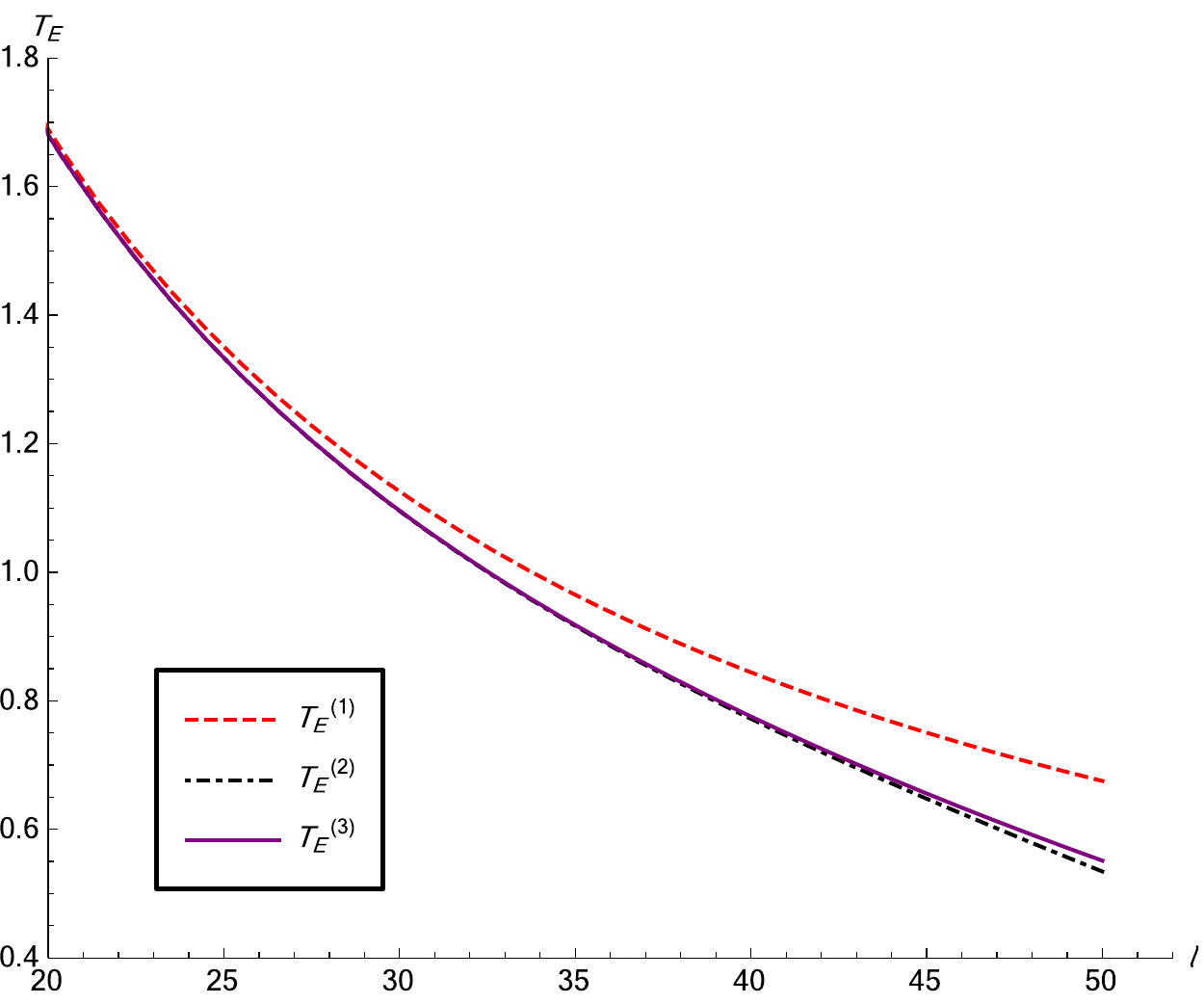}
		\caption{AdS$_5$}
	\end{subfigure}
	\hfill
	\begin{subfigure}{0.48\textwidth}
		\centering
		\includegraphics[width=\textwidth]{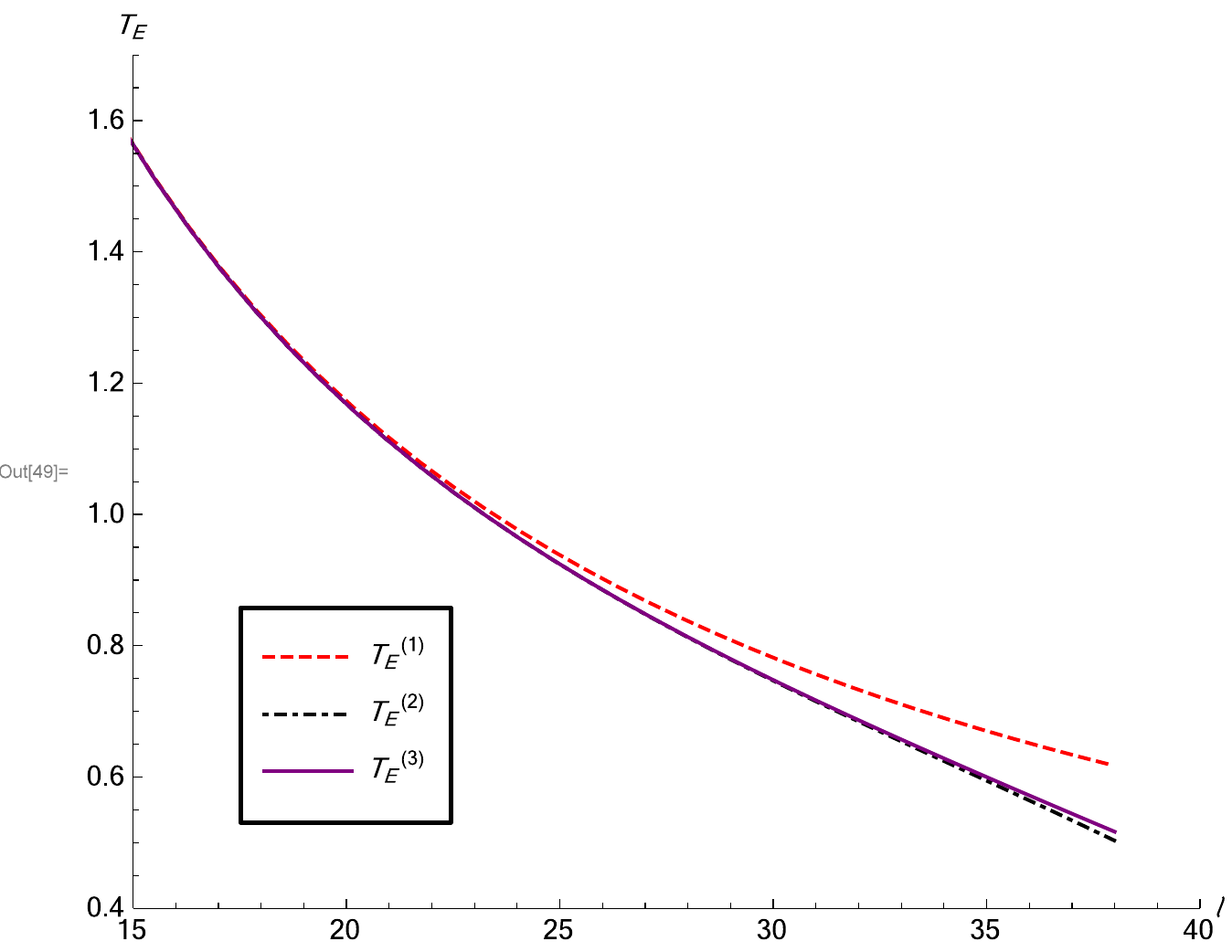}
		\caption{AdS$_7$}
	\end{subfigure}
	\caption{The entanglement temperature with and without higher order corrections of AdS plane wave for $d=4~\text{and}~6$, plot drawn for $z_I=100$.}
	\label{fig:temperature-PW}
\end{figure}
The entanglement temperatures for AdS plane wave in $5$ and $7$ dimensions is shown in figure \ref{fig:temperature-PW}, we note that in contrast with the AdS black hole, the first order correction to the temperature is negative and second order is positive. The entanglement temperature flows towards zero with increase in system size, which is expected since the background is dual to zero temperature CFT excitation. We learn that the higher order corrections are significant for any finite value of the system size $\ell$; for $\ell \to 0$, $T_E \to \infty$ and that is trivial indication of having no subsystem at all.
\section{Extension to non-conformal Dp branes}\label{sec6}
The gauge/gravity duality dictionary can also be extended to non-conformal field theories on the boundary. The supergravity description is obtained from type II solution describing $N$ coincident Dp-branes in the field theory limit \cite{Itzhaki:1998dd}. A non-conformal black p-brane background in the Einstein frame dual to a boosted field theory at finite temperature can be written as
\begin{multline}\label{dpmetric}
		ds^2 = \frac{\left(2\pi\right)^{\frac{p-2}{2}}}{g_{YM}}
		\frac{z^\frac{\theta}{4}}{r_p^{\frac{(7-p)^2}{8}}}
		\left[-\frac{f\left(z\right)}{K\left(z\right)}\frac{dt^2}{z^2} +
		K\left(z\right)\frac{\left(dy-\omega\right)^2}{z^2} +
		\frac{dx_2^2+\cdots+dx_{p}^2}{z^2} \right. \\ \left. +
		\frac{4}{(5-p)^2}r_p^{7-p}\frac{dz^2}{f\left(z\right)\,z^2} +
		r_p^{7-p}d\Omega_{8-p}^2 \right],
\end{multline}
with the dilaton being given by $e^\phi = \left(2\pi\right)^{2-p}g^2_{YM}\left(r_p^{7-p}z^{\frac{2(7-p)}{(5-p)}}\right)^{\frac{3-p}{4}}$ and the functions
\begin{equation}
	f(z) = 1 - \frac{z^{\tilde{p}}}{z_h^{\tilde{p}}}\,, ~~~K(z) = 1 + \beta^2\gamma^2\frac{z^{\tilde{p}}}{z_h^{\tilde{p}}}\,, ~~~\omega(z) = \beta^{-1}\left(1-\frac{1}{K}\right)dt,
\end{equation}
where $\tilde{p} = \frac{14-2p}{5-p}~\text{and}~\theta = - \frac{(3-p)^2}{5-p}$, the parameter $r_p$ is related to the gauge coupling by $r_p = g^2_{YM}d_pN\,.$

The space-time \eqref{dpmetric} is only conformally $AdS_{p+2} \times S^{8-p}$, for $p=3$ the conformal factor vanishes and one obtains the well known $AdS_5 \times S^5$ geometry which corresponds to a stack of $N$ D3-branes in the decoupling limit and is dual to an $\mathcal{N}=4$ SYM on the AdS boundary \cite{Maldacena:1997re}. 

We can use the Ryu-Takayanagi formula and find the HEE for non-conformal geometries using the Einstein frame metric \eqref{dpmetric}, see \cite{Ryu:2006ef, Pang:2013lpa, Mishra:2016yor, Lala:2020lcp}. For a strip of width $\ell$ as considered before: $\{-\frac{\ell}{2} \leq x_2 \leq \frac{\ell}{2};~x_3 = x_4 = \cdots = x_p = L\}$, we obtain the following area integral
\begin{align}\label{ncareaint}
	\mathcal{A}_{\gamma} &= \frac{(2\pi)^{2p-4}V_{p-1}\Omega_{8-p}}{g_{YM}^4r_p^{(7-p)}} \int_{\epsilon}^{z_*}\frac{dz}{z^{p-\theta}}\sqrt{K(z)}\sqrt{x_2^{\prime2}+\frac{4}{(5-p)^2}r_p^{7-p}f^{-1}(z)}\,,\nonumber \\
	&= \frac{2(2\pi)^{2p-4}}{(5-p)g_{YM}^4r_p^{\frac{7-p}{2}}}V_{p-1}\Omega_{8-p}\int_{\epsilon}^{z_*}\frac{dz}{z^{\tilde{p}-1}}\sqrt{K(z)}\sqrt{\bar{x}_2^{\prime2}+f^{-1}(z)}\,,
\end{align}
where we did a rescaling $\bar{x}_2(z) = \frac{(5-p)}{2}r_p^{-(7-p)}\,x_2(z)$ to reach the last line. Here, $V_{p-1} = 2\pi r_y L^{p-2}$ is the volume of the unconstrained space directions on the brane $\left(x_3, x_4, \ldots, x_p \right)$ and the y-circle, and $\Omega_{8-p}$ denotes the volume of the $(8-p)$-sphere.

The area integral is identical to \eqref{Aintegral} for the conformal case, except for the replacement $d \to \tilde{p}$. Consequently, we can use all results from previous sections by making the replacement in $d$. Instead of repeating them, we go straight into writing a first law for non-conformal cases; the components of the conserved boundary stress-energy tensor for this geometry are \cite{Mishra:2016yor}
%
%
\begin{align}
		E &= \frac{Q_pV_{p-1}\Omega_{8-p}}{16\pi G_N^{10}}\left(\frac{\tilde{p}-1}{\tilde{p}} + k\right)\frac{(7-p)\,\ell}{z_h^{\tilde{p}}}\,, \\
		P &= \frac{r_yQ_p\Omega_{8-p}}{8 G_N^{10}}\frac{(7-p)}{z_h^{\tilde{p}}}\,,\\
		N &= \frac{r_yQ_pV_{p-1}\Omega_{8-p}}{16\pi G_N^{10}}\beta\gamma^2\frac{(7-p)\,\ell}{z_h^{\tilde{p}}}\,,
\end{align}
where $Q_p$ stands for the numerical pre-factor in \eqref{ncareaint}, $Q_p = \frac{2(2\pi)^{2p-4}}{(5-p)g_{YM}^4r_p^{\frac{7-p}{2}}}$. We shall denote total volume of the boundary subsystem by $V = \ell V_{p-1}$. The entanglement chemical potential $\left(\mu_E\right)$ can again be derived perturbatively from equation \eqref{chempot}, with $d$ replaced by $\tilde{p}$ in all expressions.
%

\begin{figure}[t]
	\begin{subfigure}{0.48\textwidth}
		\centering
		\includegraphics[width=\textwidth]{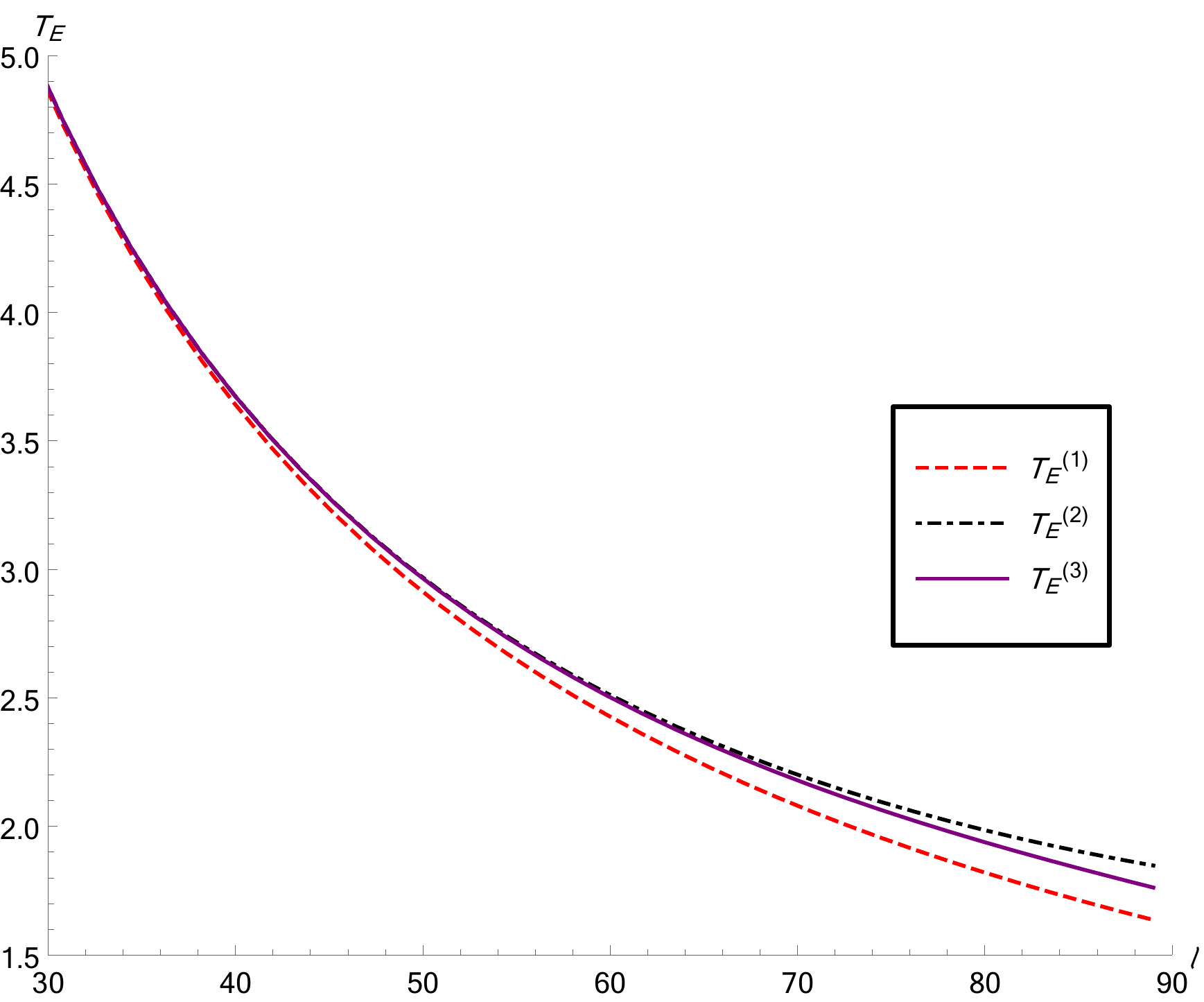}
		\caption{$p=2$}
	\end{subfigure}
	\hfill
	\begin{subfigure}{0.48\textwidth}
		\centering
		\includegraphics[width=\textwidth]{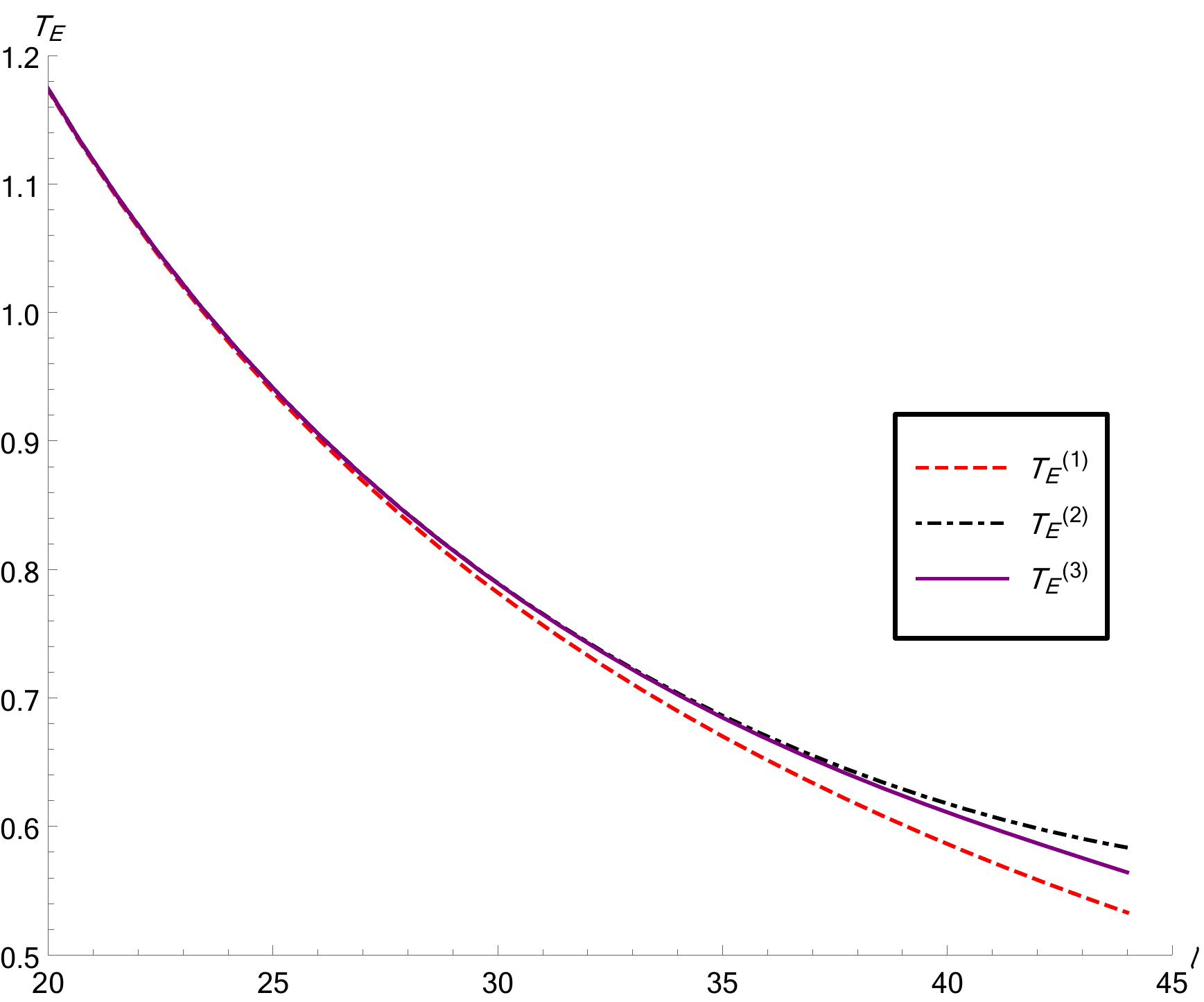}
		\caption{$p = 4$}
	\end{subfigure}
	\caption{The entanglement temperature with and without higher order corrections of D2 and D4 branes, plot drawn for $z_h=100$ and $\beta=0.25$.}
	\label{fig:temperature-Dp}
\end{figure}

One can write down a first law of entanglement involving third order corrections by following the procedure of section \ref{sec4}. The entanglement temperature at first order is
\begin{equation}
	T_E^{(1)} = \left(\frac{7-p}{\tilde{p}+1}\right)\frac{2b_0^2}{\pi b_1}\ell\,,
\end{equation}
which for $p=3$ matches with the result \eqref{temp1} of AdS$_5$; we note, however, that $\ell$ in this expression is associated with the rescaled dimension $\bar{x}_2$ from equation \eqref{ncareaint}. The higher order modifications can be determined but we do not write them explicitly, instead we refer to figure \ref{fig:temperature-Dp} for their effect on $T_E$.
\section{Conclusion}\label{sec7}
In this work, we studied the holographic entanglement entropy between a strip sub-region and its complement on the boundary of a boosted AdS black hole. We assumed the strip-width $(\ell)$ to be very narrow compared with the black hole parameter $(z_h)$ and performed a perturbation series analysis of the HEE up to third order in the dimensionless parameter $\frac{\ell}{z_h}$. Further, we tried to establish a first law of entanglement thermodynamics including corrections at this order. Due to boost in the direction $(y)$ being compact, the momentum in $y$-direction is quantized and appears as a conserved $U(1)$ charge in a lower dimensional theory. We find that it lends important contribution to the first law of entanglement. We have shown that changes in the HEE beyond leading order can be absorbed in a redefined entanglement temperature $\left(T_E\right)$ and chemical potential $\left(\mu_E\right)$, thus allowing the canonical first law relationship
\begin{equation}\label{FLEE}
	\delta S_E^{(3)} = \beta_E^{(3)}\left(\delta E - \frac{d-1}{d+1}V\delta P - \bar{\mu}_E^{(2)} \delta N \right),
\end{equation}
to remain true irrespective of the order of perturbation series. The same trick is shown to work for other holographic gravity duals e.g. AdS plane wave and non-conformal black Dp-branes as well. While we have only considered strip entangling region and third order corrections, the algorithm should work for all entangling regions and can incorporate corrections at any arbitrary order. At the same time, this way of rewriting the first law of entanglement avoids inclusion of any foreign quantity other than the conserved charges of the theory in the law.

The entanglement entropy in a general state is also known to obey a first law of entanglement with respect to the modular Hamiltonian \cite{Blanco:2013joa}
\begin{equation}\label{ModH}
	\Delta S \leq \langle \Delta H \rangle.
\end{equation}
In general, $H$ is a non-local operator; so it is curious that $H$ can somehow be expressed in terms of the subregion thermodynamic variables so that the more conventional looking eq. \eqref{FLEE} is satisfied. The resolution could be that for the special case of interest as ours, the modular Hamiltonian is local; unfortunately this suspicion cannot be confirmed unless a closed form expression for $H$ is found for a strip entangling surface in a CFT\footnote{There indeed are cases where $H$ is local and can be expressed as integral over components of the stress-energy tensor, e.g. the vacuum state of a CFT and spherical entangling surface \cite{Blanco:2013joa}.}. Another possibility is that the modular Hamiltonian effectively looks local in the small subsystem size limit $\left(\frac{\ell}{z_h} \ll 1\right)$ that we assume in this work and can be expressed in terms of subregion thermodynamic variables. In fact, it was shown in \cite{Garrison:2015lva} that the reduced density matrix always thermalizes as long as the subsystem size is much less than the total system size; this could as well be the reason that $H$ becomes proportional to subregion energy, pressure etc. and a local first law \eqref{FLEE} is satisfied\footnote{We thank the anonymous referee for their comments in this regard.}.  

While certainly not conclusive, we hope our work helps in deciding whether a first law like relationship is a generic feature of entanglement in holographic field theories.

\section*{Acknowledgements}
SM acknowledgess the Yukawa Institute for Theoretical Physics at Kyoto University, where a version of this work was presented during the YITP workshop YITP-W-20-03 on `Strings and Fields 2020'.
\appendix
\section{List of integrals}\label{append1}
In the following, we list the integrals encountered while solving the equations \eqref{turnpt} and \eqref{RTintegral} and write their simplified form in terms of Beta functions, first let us recall the definition
\begin{equation}
	\int_{0}^{1}\frac{dy\,y^{nd-1}}{\sqrt{R}} = \frac{B\left(\frac{nd}{2d-2},\frac{1}{2}\right)}{2d-2}\equiv b_{n-1}\,,
\end{equation}
where $R \equiv 1-y^{2d-2}$.

\noindent We start with the integrals occurring while solving for the turning point $\left(z_*\right)$ from equation \eqref{turnpt}
\begin{align}
	\int_0^1 \frac{y^{d-1}\left(1-y^d\right)dy}{\sqrt{R}} &= \frac{d+1}{d-1}b_1 - \frac{1}{d-1}b_0\,,\\
	\int_0^1 \frac{y^{2d-1}\left(1-y^d\right)dy}{R^{3/2}} &= - \frac{2d+1}{d-1}b_2 - \frac{d+1}{d-1}b_1\,,\\
	\int_0^1 \frac{y^{d-1}\left(1-y^d\right)^2dy}{R^{5/2}} &= \frac{2d+1}{(d-1)^2}b_2 + \frac{8}{3}\frac{(d+1)(d-3)}{(2d-2)^2}b_1 - \frac{4}{3}\frac{2d-3}{(2d-2)^2}b_0\,,\\
	\int_0^1 \frac{y^{3d-1}\left(1-y^d\right)dy}{R^{3/2}} &= \frac{3d+1}{d-1}b_3 - \frac{2d+1}{d-1}b_2\,,\\
	\int_0^1 \frac{y^{2d-1}\left(1-y^d\right)^2dy}{R^{5/2}} &= \frac{4}{3}\frac{(3d+1)(d+3)}{(2d-2)^2}b_3 - \frac{2(2d+1)}{(d-1)^2}b_2 -\frac{4}{3}\frac{(d+1)(d-3)}{(2d-2)^2}b_1\,,\\
	\begin{split}
		\int_0^1 \frac{y^{d-1}\left(1-y^d\right)^3dy}{R^{7/2}} &= -\frac{8}{15}\frac{(d-5)(d+3)(3d+1)}{(2d-2)^3}b_3 + \frac{24}{5}\frac{(2d-5)(2d+1)}{(2d-2)^3}b_2\\ &~~~~+ \frac{8}{5}\frac{(3d-5)(d-3)(d+1)}{(2d-2)^3}b_1 - \frac{8}{15}\frac{(4d-5)(2d-3)}{(2d-2)^3}b_0\,.
	\end{split}
\end{align}
Next, The integrals that arise when solving for the extremized area are
\begin{align}
	&\int_{0}^{1}\frac{dy}{y^{d-1}\sqrt{R}} = -\frac{b_0}{d-2}\,,\\
	&\int_{0}^{1}\frac{y^ddy}{y^{d-1}\sqrt{R}} = \frac{d+1}{2}b_1\,,\\
	&\int_{0}^{1}\frac{y^{2d-2}(1-y^d)dy}{y^{d-1}R^{3/2}} = \frac{d+1}{d-1}b_1 - \frac{1}{d-1}b_0\,,\\
	%
	&\int_{0}^{1}\frac{y^{2d}dy}{y^{d-1}\sqrt{R}} = \frac{2d+1}{d+2}b_2\,,\\
	&\int_{0}^{1}\frac{y^{3d-2}(1-y^d)dy}{y^{d-1}R^{3/2}} = \frac{2d+1}{d-1}b_2 - \frac{d+1}{d-1}b_1\,,\\
	&\int_{0}^{1}\frac{y^{4d-4}(1-y^d)^2dy}{y^{d-1}R^{5/2}} = \frac{4}{3}\frac{3d(2d+1)}{(2d-2)^2}b_2 - \frac{8}{3}\frac{2d(d+1)}{(2d-2)^2}b_1 + \frac{4}{3}\frac{d}{(2d-2)^2}b_0\,,\\
	&\int_{0}^{1}\frac{dyy^{3d}}{y^{d-1}\sqrt{R}} = \frac{3d+1}{2d+2}b_3\,,\\
	&\int_{0}^{1}\frac{y^{4d-2}(1-y^d)dy}{y^{d-1}R^{3/2}} = \frac{3d+1}{d-1}b_3 - \frac{2d+1}{d-1}b_2\,,\\
	&\int_{0}^{1}\frac{y^{5d-4}(1-y^d)^2dy}{y^{d-1}R^{5/2}} = \frac{4}{3}\frac{4d(3d+1)}{(2d-2)^2}b_3 - \frac{8}{3}\frac{3d(2d+1)}{(2d-2)^2}b_2 + \frac{4}{3}\frac{2d(d+1)}{(2d-2)^2}b_1\,,
\end{align}
\begin{equation}
	\begin{split}
		\int_{0}^{1}\frac{y^{6d-6}(1-y^d)^3dy}{y^{d-1}R^{7/2}} &= - \frac{8}{15}\frac{4d(6d-2)(3d+1)}{(2d-2)^3}b_3 - \frac{8}{5}\frac{3d(5d-2)(2d+1)}{(2d-2)^3}b_2 \\ &~~~~+ \frac{8}{5}\frac{2d(4d-2)(d+1)}{(2d-2)^3}b_1 - \frac{8}{15}\frac{d(3d-2)}{(2d-2)^3}b_0\,.
	\end{split}
\end{equation}
\bibliographystyle{JHEP}
\bibliography{references}
\end{document}